\documentstyle[11pt,psfig]{article}

\newcommand{\prob}{\mbox{Prob}}
\newcommand{\bee}{\begin{equation}}
\newcommand{\eee}{\end{equation}}
\newcommand{\ble}{\begin{equation}}
\newcommand{\ele}{\end{equation}}
\newcommand{\mper}{\mbox{\ .}}
\newcommand{\mcom}{\mbox{\ ,}}
\newcommand{\tr}{{\rm tr}}
\newcommand{\del}{\partial}

\newcommand{\laur}{\marginpar{\raisebox{3in}{\small \tt LAUR 91-3331}}}
\title{Zeta function for the Lyapunov exponent of a product of
random matrices}
\author{Ronnie Mainieri \\
{\it Niels Bohr Institute} \\
{\it Blegdamsvej 17, Copenhagen \O, 2100 Denmark }
 \\
 {\it and} \\
 {\it Center for Nonlinear Studies} \\
 {\it Los Alamos National Laboratory,
 Los Alamos, NM 87545\thanks{current address}} \\
 {\small \tt  ronnie@goshawk.lanl.gov}
}

\date{PACS 05.50.+q 05.45.+b 02.50.Ga}

\begin{document}

\maketitle
\laur

\begin{abstract}
A cycle expansion for the Lyapunov exponent of a product of random
matrices is derived.  The formula is non-perturbative and numerically
effective, which allows the Lyapunov exponent to be computed to high
accuracy.  In particular, the free energy and the heat capacity
are computed for the one-dimensional
Ising model with quenched disorder.  The formula is
derived by using a Bernoulli dynamical system to mimic the
randomness. 
\end{abstract}

The product of random matrices often appears in the study of disordered
materials and of dynamical systems.  The physical quantities of these
systems are related to the rate of growth of the random product --- the
Lyapunov exponent.  For example, in the study of an Ising model with
quenched randomness the Lyapunov exponent is proportional to the free
energy per particle; in the Schr\"{o}dinger equation with a random
potential, the Lyapunov exponent is proportional to the localization
length of the wave function; and in the motion of a classical particle,
the Lyapunov exponent indicates the degree of sensitivity to initial
conditions (chaos).  Since Dyson \cite{DysonOscillators} studied a
system of harmonic oscillators with random couplings, many problems
have been reduced to the study of a Lyapunov exponent.  In these
problems the product of random matrices appears when a discrete version
of a differential operator is considered, or when the problem is solved
on a lattice.   Further applications of Lyapunov exponents, and related
derivations, are reviewed in the paper by Alexander {\it et al.\/}
\cite{AlexanderReview} and the book by Crisanti, Paladin and Vulpiani
\cite{PaladinBook}.

Few are the analytic results for the Lyapunov exponent of a product of
random matrices, and very little has been determined about related
systems without resorting to Monte Carlo simulations.  A theorem of
Oseledec \cite{OseledecThm} states that the norm of the random product
grows exponentially with the number of multiplied terms at a rate given
by the Lyapunov exponent, but the theorem does not provide a method for
determining the exponent.  The two known methods for calculating the
Lyapunov exponent, weak disorder expansions
\cite{DerridaTwo,DerridaWeak,ZanonDegenerate} and microcanonical
approximations \cite{PaladinMicro}, have limitations.  The weak
disorder expansion imposes conditions on the eigenvalues of the
matrices and is difficult to carry out to high orders; and the
microcanonical method, while general, does not provide a systematic
expansion and is difficult to apply to large matrices.  In this Letter
a formula for computing the Lyapunov exponent will be derived.  It is
simple to evaluate and is non-perturbative in character, with the first
few terms providing a good numerical approximation.  In particular, it
gives all thermodynamic quantities for the one-dimensional Ising model
with a discrete valued random magnetic field when the disorder
averaging is done over the free energy (this is the more difficult
quenched disorder case).

The formula is obtained by viewing the random product as a statistical
mechanical system which is solved using the cycle expansion
\cite{CvitanovicCycles} of its thermodynamical zeta function
\cite{RuelleZeta}.  Cycle expansions have been very successful for
obtaining non-perturbative expansions of chaotic dynamical systems
\cite{Recycle1,Recycle2}, of generalized Ising systems
\cite{Thermodynamic,thesis}, and of scattering problems in quantum
mechanics \cite{CvitanovicQuantization}.

We will consider the product 
\bee
	G^{(n)} = \prod_{0<k\leq n} T_k
\eee
of matrices $T_k$ chosen at random from a discrete set.  This includes
many cases of interest and can be used to approximate a continuous
distribution.  The maximal Lyapunov exponent $\gamma$ can be
expressed \cite{MargulisReview} as the rate of exponential
growth of the norm of the product $G^{(n)}$ with the number $n$ of
matrices multiplied:
\bee
        \gamma = \lim_{n \rightarrow \infty} \frac{1}{n}
                 \langle \ln \| G^{(n)} \| \rangle
\eee
The average is over all possible realizations of the product, each
product taken with the appropriate probability.  The theorem of Oseledec
\cite{OseledecThm} guarantees that the limit exists for almost every
realization.  The definition of
the Lyapunov exponent appears to depend on the matrix norm chosen, but
it can be shown \cite{MargulisReview} that its value remains unchanged
as long as equivalent norms are used.  For the finite dimensional
vector space of $n \times n$ matrices all norms are equivalent, making
the Lyapunov exponent independent of the norm chosen.

Because it is very difficult to handle the logarithm inside the
average we will substitute a power and a derivative for the
logarithm,  and write the Lyapunov exponent as
\bee
        \gamma = 
                \left.
                  \lim_{n \rightarrow \infty} 
		  \del_\alpha
                  \frac{1}{n}
                  \langle \|G^{(n)}\| ^{\alpha} \rangle
                \right|_{\alpha = 0}
        \mper
\eee
To determine the averages, introduce the generating function 
\bee
	\zeta(z,\alpha) = \exp \left(
			      \sum_{n\geq 1} 
			      \frac{z^n}{n}
			      \langle \| G^{(n)} \|^\alpha \rangle
			\right)
	\mcom
\eee
which is the Ruelle zeta function \cite{Thermodynamic} for a
statistical mechanical system with $\langle\|G^{(n)}\|^\alpha\rangle$
as the free energy.  The zero $\hat{z}(\alpha)$ of
$1/\zeta(z,\alpha)$ gives the exponential growth of
\begin{math}
	\langle \|G^{(n)}\|^\alpha \rangle
\end{math}
(see Ref.\ \cite[Thm 5.29]{Thermodynamic}), and by using the special
values $\hat{z}(0)=1$ and $\del_{z}\zeta^{-1}(1,0)=-1$, the Lyapunov
exponent can be re-expressed as
\ble
        \gamma =
	- \del_\alpha \ln \hat{z}(\alpha = 0) =
	- \del_{\alpha}\zeta^{-1}(1,0)
        \mper
	\label{deriv}
\ele

The expression in terms of the
derivative of the zeta function is of no advantage unless it can
be computed in an efficient manner.  If the terms of the zeta
function satisfy certain combinatorial properties the inverse
zeta function can be written as a cycle expansion
\cite{CvitanovicCycles}, which is
rapidly convergent and offers a practical scheme for evaluating
the Lyapunov exponent.
The average $z^n\langle\|G^{(n)}\|\rangle$
is the sum of terms of the form
\bee
        t_{G} = z^{n(G)} \prob(G) \|G\|^{\alpha}
\eee
with $n(G)$ being the number of matrices in the product $G$. 
If the weights are cyclic, as in 
\begin{math}
        t_{AAB} = t_{ABA} = t_{BAA}
\end{math},
and multiplicative, as in
\begin{math}
        t_{ABAB} = (t_{AB})^2
        \mcom
\end{math}
then the inverse zeta function can be expanded into a cycle
expansion \cite{Recycle1,thesis},
\bee
        \zeta^{-1}(z,\alpha) = \prod_{G \in P} (1-t_G)
\eee
with the product being over the set $P$ of all possible prime
products.  A product of matrices is prime if it is not the repeat of a
smaller length product. Two products are equivalent for the expansion
if they differ by a cyclic rotation.  For example, if $AB$ is in the
set $P$, then $BA$ does not need to be in the set as it is equivalent
to $AB$ by cyclic rotation.  Also $ABAB$ and $BABA$ need not be in $P$
as they are repeats of $AB$ or of its cyclic rotation.

To continue the derivation we use the independence of the
Lyapunov exponent on the norm, and choose the most convenient
norm for the cycle expansion.  If we choose the norm absolute
value of the largest eigenvalue (or eigenvalues), and write it in the
peculiar form
\bee
        \|G\|^\alpha = \lim_{n \rightarrow \infty} 
                | \tr G^n |^{\alpha/n}
\eee
then it is simple to verify that the weights $t_G$ are
multiplicative and cyclic.  They are multiplicative because
\bee
        \|G^p\|^\alpha = \lim_{n \rightarrow \infty}
                | \tr G^{pn} |^{\alpha/n} =
        \left( \lim_{pn \rightarrow \infty} 
                | \tr G^{pn} |^{1/(pn)} \right)^{\alpha p}  =
        \| G \|^{\alpha p}
        \mper
\eee
And they are cyclic because the trace is cyclic.  The $\prob(\cdot)$
part of the weight is multiplicative and cyclic, as the product of
numbers is.  The cycle expansion for $\langle\|G^{(n)}\|^\alpha\rangle$
is then obtained by expanding the infinite product
\ble
        \zeta^{-1}(z,\alpha) = \prod_{G \in P} \left( 1 -
        z^{n(G)} \prob(G) \|G\|^{\alpha} \right)
	\label{cycle}
\ele
into a power series in $z$.  This is essential when computing the zeros
of $\zeta^{-1}$.  The power series in $z$ converges as long as the
matrices are hyperbolic (not all eigenvalues equal to one in
modulus).  The cycle expansion (\ref{cycle}) can be used to compute the
Lyapunov exponent in equation (\ref{deriv}).  In the case that there
are two matrices forming the random product, $A$ with probability $p$
and $B$ with probability $q=1-p$, the first few terms of the expansion
of (\ref{deriv}) are:
\bee
    \begin{array}{rcl}
        - \gamma & = & 
        p \ln \| A\| + q \ln \|B \| + 
%	\mbox{} \\ & & \mbox{} +
        pq ( \ln \| AB \| - \ln \| A \| - \ln \|B\| ) + \mbox{} \\
        & & \mbox{} +
        ppq ( \ln \| AAB \| - \ln \|AB \| - \ln \|A \| ) + \mbox{} \\
        & & \mbox{} +
        pqq ( \ln \| ABB \| - \ln \|AB \| - \ln \|B \| ) +
        \cdots
    \end{array}
\eee
General expressions and a more detailed derivation can be found
in Ref.\ \cite{comming}.

\begin{table}
  \begin{center}
	\begin{tabular}{lll}
	method & Lyapunov & heat capacity \\
	\hline
	Weak Disorder &  $1.5$ & --- \\
	Microcanonical & $1.20$ & --- \\
	Monte Carlo & $1.1773$ & $0.4$ \\
	Zeta ($n=15$) & $1.17727361334268$ & $0.3664$ 
	\end{tabular}
  \end{center}
  \caption[boo]{The Lyapunov exponent and its second derivative
	(proportional to the heat capacity) for the transfer matrices
	(see equation \protect\ref{matrix}) when
	$j=0.3$ and $h=1.4$. The Monte Carlo and zeta function
	values are quoted so that errors are
	limited to the last digit.}
  \label{methods}
\end{table}

One test of the expansion for the Lyapunov exponent is a product of
random matrices that appears in the study of the one-dimensional Ising
model with coupling constant $J$ and a random magnetic field assuming
the values $\pm h$.  The two matrices, chosen with equal probability
and after factoring out a common term, are of the form
\ble
	\left[
	\begin{array}{cc}
	 1 & ab^{k} \\
	 a & b^{k}
	\end{array}
	\right]
	\label{matrix}
\ele
where $k$ is $+1$ or $-1$, $a$ is $\exp(-2J)$, and $b$ is
$\exp(-2h)$.
In this case both matrices have eigenvalues that are
real and different from one, and are therefore hyperbolic.  The
Lyapunov exponent for these matrices can also be computed by Monte
Carlo simulations, weak disorder expansions, and microcanonical
approximations.  Table \ref{methods} has the results for these
methods.  The Monte Carlo calculations corresponds to 128 realizations of
a product of $10^6$ matrices; the weak disorder expansion is carried to
fifth order; there is no
error estimate for the microcanonical method, except for a rigorous
upper bound; and the value for the zeta function method was
computed by including cycles up to length fifteen.  The zeta
function expansion takes 17 seconds on a Sparcstation 1 computer,
whereas if the Monte Carlo simulation had to be carried out to
the precision obtained in the cycle expansion, it would require several
hundred years of Sparcstation 1 time.  The weak disorder expansion
could in principle match the accuracy of the zeta function expansion if
the terms of higher order were known.  Also in table \ref{methods} the
second derivative of the Lyapunov exponent (proportional to the heat
capacity at $\beta = 1$) is computed by numerical differentiation.
Notice that, except for the zeta function, all analytic methods fail to
provide a value for the second derivative and that even though Monte
Carlo does estimate the derivative with one digit, a better estimate
would require a prohibitive amount of computer time.

\begin{figure}
	\centerline{\psfig{file=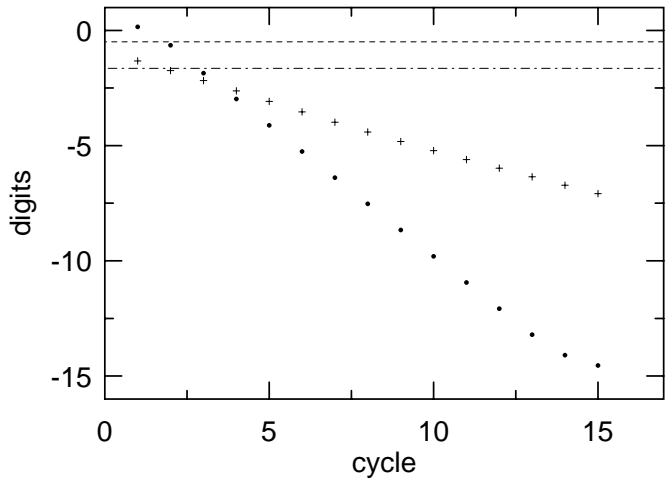,height=2.5in}}
	\caption{Number of digits that remain constant as longer cycles
	are included in the expansion of the Lyapunov exponent.
	The bullets ($\bullet$) are for the random Ising model
	and the crosses ($+$) are for the
	degenerate $3\times 3$ matrices.  Also indicated in the
	plot are the accuracy of the weak disorder expansion (short
	dashes) and the microcanonical approximation (long dashes).}
	\label{digits}
\end{figure}

To study the convergence of the zeta function method one can plot
the rate at which digits are gained in the value of the
Lyapunov exponent as longer cycles are included in the expansion.
The better a system is understood, the better the nature of
the convergence.
In figure \ref{digits} the number of correct digits as a function
of the largest cycle considered is plotted.
If $\gamma_n$ is the approximation to the Lyapunov exponent when cycles up to
length $n$ are included, then the number of digits is defined as
$d(n)=\log_{10}(\gamma_{n-1}-\gamma_{n})$.  The straight line indicates
that the convergence is exponentially fast in the length of the product.

\begin{figure}
	\centerline{\psfig{file=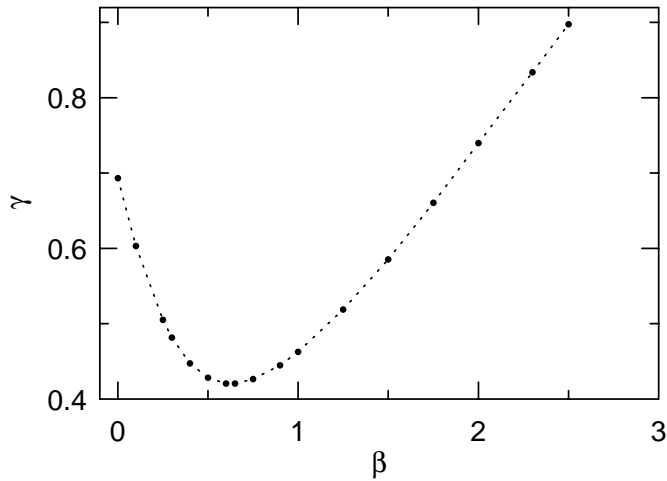,height=2.5in}}
	\caption[boo]{ Lyapunov exponent (free energy) for the
	pair of matrices (\protect\ref{matrix}) that describe the Ising
	model with quenched randomness as a function of the
	inverse temperature $\beta$.
	The dotted line is an interpolation of the computed points.}
	\label{FigPhase}
\end{figure}

To further illustrate the method, figure \ref{FigPhase} has a plot of
the Lyapunov exponent for the Ising model pair of matrices chosen with
equal probability and in units where $J=h=\beta$.  In the plot all
points can be computed to machine precision, and the convergence rate
is similar to that of figure \ref{digits}.  Thus the method is not
limited to small values of the inverse temperature $\beta$ as the weak
disorder expansion, and can be used to obtain thermodynamic quantities
at any temperature.

The weak disorder expansion cannot be applied when there are repeated
eigenvalues, so to illustrate the zeta function method the Lyapunov
exponent has been computed for a pair of matrices with degenerate
eigenvalues.  The random products are formed from a pair of three by
three matrices, which have the same eigenvalues, do not commute, and
are not related by a similarity transformation.  The eigenvalues are 2,
2, and 1.  The exponential convergence of the method is not affected by
the largest eigenvalue being degenerate, nor by the presence of an
indifferent eigenvalue (the one of value one).  In figure \ref{digits}
the crosses are the plot of the number of non-changing digits of the
Lyapunov exponent as a function of the cycle length.

The cycle expansion developed in this Letter for the Lyapunov exponent
is an efficient computational tool.  It can be applied to a wide
variety of matrix products without the limitations of other methods,
and excludes only the matrices where all the eigenvalues are one in
modulus.  The method has been successfully applied to Ising models with
a random magnetic field on a strip \cite{comming}, and also in
reproducing the branch point at zero temperature predicted by Derrida
and Hilhorst \cite{DerridaSingular} in the same model.

I would like to thank the hospitality of the Neils Bohr Institute
where this worked was carried out under the financial support of
the {\sc NATO/NSF} post-doctoral fellowship RCD-9050092.  It is
also a pleasure to acknowledge discussions with
Predrag Cvitanovi\'{c},
Robert Ecke,
Marco Isopi,
Giovanni Paladin,
and
Ruben Zeitak.

\bibliographystyle{unsrt}

\end{document}